# Direction dependent thermoelectric properties of layered compound $In_2Te_5$ single crystal

*Anup V. Sanchela*[a,b*], *Ajay D. Thakur*[c,d], *C.V. Tomy*[b]

[a]*Department of Physics, Pandit Deendayal Energy University, Raysan, Gandhinagar 382007, India*
[b]*Department of Physics, Indian Institute of Technology Bombay, Powai, Mumbai 400076, India*
[c]*School of Basic Sciences, Indian Institute of Technology Patna, Bihta 801118, India*
[d]*Centre for Energy & Environment, Indian Institute of Technology Patna, Bihta 801118, India*

**Abstract.** We analyze the anisotropic electrical and thermal transport measurements in single crystals of $In_2Te_5$ belonging to monoclinic space group C12/c1 with the temperature gradient applied || and ⊥ to the crystallographic *c*-axis of the crystals. The thermal conductivity along the c-axis ($\kappa_{||c}$) was found to smaller by a factor of 2 compared to the thermal conductivity along the direction perpendicular to the *c*-axis ($\kappa_{\perp c}$) over the entire temperature range. In contrast, the Seebeck coefficient along the *c*-axis ($S_{||c}$) was found to be higher than its value along the direction perpendicular to the *c*-axis ($S_{\perp c}$). At the room temperature, the figure of merit $zT_{||c}$ is found to be 4 times larger as compare to the figure of merit $zT_{\perp c}$.

[*]Corresponding Author
Email: anup.sanchela@sot.pdpu.ac.in

## 1. Introduction

Over the past decade, thermoelectric materials have attracted considerable research interest because of their possible applications in direct solid state energy conversion between waste heat and power generation (using thermoelectric generators) as well as for the purpose of electronic refrigeration [1-4]. The performance of thermoelectric materials can be quantified by the dimensionless figure of merit $zT= S^2T/\rho\ (\kappa_l + \kappa_e)$. Here $S$ is the thermopower, $\rho$ is the electrical resistivity, $\kappa_l$ is the lattice thermal conductivity, $\kappa_e$ is the electronic thermal conductivity and $T$ is the absolute temperature such that a large $zT$ value is desirable for improving the efficiency of thermoelectric generators and coolers. Over the past few years, several strategies were applied to reduce the phonon thermal conductivity, such as disordering the system by introducing point defects [5], creating resonant scattering by localized rattling atoms [4, 6-8], creating maximum interfaces by ball milling and hot pressing or spark plasma sintering (SPS) [4, 6, 9], introducing nanoscale precipitates in thermoelectric materials [10], synthesized the nano sheets [11] etc. Utilizing the aforementioned techniques, it is found to be possible to increase the phonon scattering thereby minimizing lattice thermal conductivity.

Anisotropy is another important parameter which plays a crucial role in enhancing the $zT$ values in thermoelectric materials. Layered compounds are the natural choice for exploiting this effect and hence the direction dependent thermoelectric properties have been investigated in detail in many layered compounds. A few of these layered compounds, which are considered as high performance layered thermoelectric materials, are $CsBi_4Te_6$, $Ag_2Te$, $Bi_2Te_3$ and SnSe. $CsBi_4Te_6$ crystallizes in a monoclinic structure (space group $C2/m$) and has lattice parameters, $a$ = 51.9 Å, $b$ = 4.4 Å and $c$ = 14.5 Å. The crystal structure comprises of alternating slabs of $Bi_4Te_6$, separated by layered $Cs^+$ ions (see Fig. 1(a)). This type of unique structure prevents the motion of $Cs^+$ atoms along the perpendicular crystallographic *b*-axis. These localized $Cs^+$ ions behave like rattlers and the resonant scattering of phonons from these rattlers reduces the lattice thermal conductivity. The Bi-Bi bonds interconnecting the slabs of $Bi_4Te_6$ are responsible for the anisotropy of charge transport and the small effective mass along the crystallographic *b*-axis and also for the narrow energy gap (~0.04 eV). Because of the narrow band gap and a low lattice thermal conductivity, the *zT* value reaches ~0.8 at 225 K in doped $CsBi_4Te_6$ [6, 12, 13].

$Bi_2Te_3$ is rhombohedral (space group $R\overline{3}m$) with lattice parameters, *a* = 4.38 Å and *b*= 30.50 Å [14]. The crystal structure consists of atomic sheets in the *a*-*b* plane with Te1-Bi-Te2-Bi-Te1 quintets (see Fig. 1(b)). These sheets are held together only by a weak van der Waals force along the *c*-axis. Therefore $Bi_2Te_3$ possesses a structural anisotropy. B. Poudel et al. [9] reported a maximum *zT* value of 1.4 at 373K in *p*-type $Bi_{2-x}Sb_xTe_3$ samples. X. Yan et al. [15] reported *zT* values in the range of 0.85 to 1.04 at 398K in *n*-type $Bi_2Te_{3-x}Se_x$ samples. In both the cases, the samples were prepared by ball-milling and hot pressing technique. Using an entirely different fabrication technique, i.e., creating a superlattice structure consisting of $Bi_2Te_3/Sb_2Te_3$, an outstanding *zT* value of ~2.4 at 300K was obtained in $Bi_2Te_3$ along the direction perpendicular to the layers [12, 16].

$Ag_2Te$ crystallize in the monoclinic structure with a *P*21/*c* space group and the lattice parameters are *a* = 8.16 Å, *b* = 4.46Å and *c* = 8.97 Å. The structure consists of triple layers; top and bottom layers comprise alternating pairs of Ag and Te atoms and the middle layer consists of only Ag atoms (see Fig. 1(c)) [17]. The room temperature $\beta$-$Ag_2Te$ (monoclinic structure) phase transforms into $\alpha$-$Ag_2Te$ (FCC cubic structure) phase at 417 K. $\alpha$-$Ag_2Te$ is also known as a Ag-ion conductor since the Ag ions start hoping to the interstitial sites when the phase transition begins. This hoping not only enhances the electrical conductivity but also reduces the lattice thermal conductivity [18]. Recently, Lee et al. have found a high *zT* value of 1.5 at 700K in bulk composites $Sb_2Te_3/Ag_2Te$ [19].

Recently, Zhao et al. reported another layered compound, SnSe which exhibits an ultra low thermal conductivity and a very high *zT* of ~2.6 along the *b*-axis. Due to anisotropy, the *zT* is only 2.3 along the *c*-axis and 0.8 along the *a*-axis. SnSe forms in a layered orthorhombic structure with space group *Pnma* at room temperature. A phase transition occurs at high temperatures about 750-800 K and the space group transforms in to a high symmetry phase *Cmcm* [20]. The crystal structure of SnSe possesses eight atoms (two adjacent double layers) in one primitive cell. The unique atomic stacking and anisotropic zigzag structure (see Fig. 1(d)) are responsible for the ultra low thermal conductivity and high *zT* in SnSe [21, 22].

$In_2Te_5$ is a compound which is reported to have a natural layered structure. $In_2Te_5$ structure was solved by Walton et al. [23] and found that $In_2Te_5$ forms in two different space groups, *Cc* and *C*12/*c*1. The structure with space group *Cc* (labeled as $In_2Te_5$ (I)) results in lattice parameters $a$ = 4.39 Å, $b$ = 16.39 Å, and $c$ = 13.52Å where as the structure with space group *C*12/*c*1 ($In_2Te_5$ (II)) forms with cell parameters, $a$ = 16.66 Å, $b$ = 4.36Å and $c$ = 41.34 Å. We have prepared the $In_2Te_5$ (II) phase (*C*12/*c*1 phase) which is confirmed by the Rietveld refinement analysis [23]. We have investigated the anisotropic thermoelectric properties of this monoclinic compound (space group *C*12/*c*1), by preparing single crystals, the results of which are present in this paper. We demonstrate that by changing the thermal gradient direction, the lattice thermal conductivity and the thermopower can be improved near the room temperature, thus leading to improvement in the figure of merit *zT* [24]. These studies are relevant in the context that the group III and VI semiconducting materials are used in practical applications like developing the heterojunctions, optoelectronic components, Schottky barriers, etc [25].

## 2. Experimental details

Single crystals of $In_2Te_5$ were grown by the modified Bridgman technique. Pieces of In (99.99%) and Te shots (99.99%) were weighed in stoichiometric ratio and sealed in a quartz ampoule after evacuation ($10^{-5}$ mbar). The charge was heated up to 500 ℃ over a time period of 15 hours, kept at this temperature for 24 hours and then cooled slowly (2 ℃/h) to 470 ℃ where it was left for 24 hours [25, 26]. The charge was then furnace-cooled to room temperature. Well-grown single crystals could easily be separated from the charge for various measurements. Typical dimension of the single crystals chosen for the present study is 4.54 × 1.96 × 1.27 $mm^3$. The nominal composition was examined by the energy dispersive X-ray analysis method. The observed values of the stoichiometery were found to be within 1.5% of the starting composition. The single crystal morphology was checked with a Field Emission Scanning Electron Microscope (FESEM) and the single crystalline nature was confirmed by the High Resolution Transmission Electron Microscope (HRTEM).

## 3. Results and discussion

Figure 2(a) shows the XRD pattern of the powdered single crystals of $In_2Te_5$. Rietveld refinement was performed on the obtained pattern using the known monoclinic structure with space group *C*12/*c*1. All the observed peaks could be indexed and a good agreement between the experimental data (open circles) and the calculated pattern (solid line) suggests the formation of $In_2Te_5$ in single phase. In order to determine the orientation of the single crystals used for the present studies, the X-ray diffraction patterns were obtained from the single crystals. One such pattern is shown in Fig. 2(b). It is clear from the pattern that the crystal planes are grown along the *c* direction (only [0, 0, 6] and [0, 0, 12] reflections are seen in the measured $2\theta$ range).

We show in Fig. 3(a), the structure of $In_2Te_5$. The dark pink and the dark blue spheres represent the indium and the tellurium atoms, respectively. The structure consists of In-Te rings interconnected by a cross-linked bunch of three Te atoms along the plane. These In-Te planes are stacked along the crystallographic *c*-axis to form a unit cell. Such a structure has been seen to possess a large unit cell with very different lattice parameters [12, 23]. Figure 3(b) shows the Small Area Electron Diffraction (SAED) pattern obtained for one of the single crystals used in our measurements. Characteristic bright spots indicating single crystalline nature of the specimen is self-evident. From the HRTEM photograph (Fig. 3(c)), the lattice spacing between the two neighboring atomic planes were estimated to be about 3.38Å (marked by two parallel lines) which is very close to the reported inter planar distance of 3.39Å between the two adjacent (0,0,12) planes [JCPDS#00-031-0602]. Figure 3(d) shows the optical micrograph of a cleaved piece of a single crystal belonging to the same batch where the layered structure of the single crystal $In_2Te_5$ is clearly visible. Summary of the crystallographic data is given in Table 1.

The temperature dependence of the electrical resistivity $\rho(T)$ along the two crystallographic directions $\rho_{\parallel c}$ and $\rho_{\perp c}$ is shown in Fig. 4. Resistivity behavior is typical of this compound, as reported in ref. [25], with a maximum around 300K due to the change over of conductivity from impurity to intrinsic and the associated change in the hole mobility and carrier concentration [25]. $\rho_{\parallel c}$ and $\rho_{\perp c}$ follow different paths, indicating the anisotropic behavior. At about 300 K, $\rho_{\parallel c}$ = 0.096 Ωm and $\rho_{\perp c}$ = 0.12 Ωm.

The variation of the Seebeck coefficient $S(T)$ with respect to temperature along the two directions, $S_{\parallel c}$ and $S_{\perp c}$, for the single crystal of $In_2Te_5$ is shown in Fig. 5(a). The positive sign of the Seebeck coefficient indicates that the majority charge carriers are $p$ type (holes) throughout the temperature range. The highest value of $S_{\parallel c}$ is 480 $\mu$V/K at about 273 K and $S_{\perp c}$ is 408 $\mu$V/K at about 267 K. During the initial increase of temperature, $S(T)$ in both the orientations remain the same, but after about 200 K, $S_{\parallel c}$ increases faster as compared to $S_{\perp c}$ ($S_{\parallel c} > S_{\perp c}$). The maximum in $S(T)$ is in agreement with the maximum observed in the conductivity of this sample due to the change of conductivity from impurity to intrinsic and the variation of hole mobility and hole concentration with temperature [25].

In order to find out the energy band gap ($E_g$) for our crystal, we have used the Goldsmid and Sharp formula [27],

$$E_g = 2eS_{max}T_{max}$$

where $e$ is the hole charge, $S_{\max}$ is the maximum value of the thermopower and $T_{\max}$ is the corresponding temperature. Using the values of $S_{\max}$ and $T_{\max}$ from Fig. 5(a), we obtain $E_g$ = 0.21 eV at 267 K and 0.26 eV at 273 K for $S_{\perp c}$ and $S_{\parallel c}$ directions, respectively.

In order to measure the direction dependent thermal conductivity, temperature gradient was applied either along or perpendicular to the *c*-axis. The schematic diagram of the layered configuration with temperature gradient applied along parallel or perpendicular to the crystallographic *c*-axis is shown in the inset of Fig. 5(a). The temperature dependence of the total thermal conductivity ($\kappa$) along the two crystallographic directions ($\kappa_{\parallel c}$ and $\kappa_{\perp c}$) is shown in Fig. 5(b). Both the thermal conductivities follow the power law (the powers being 0.37 and 0.60 for $\kappa_{\parallel c}$ and $\kappa_{\perp c}$, respectively) up to 35 K with a typical phonon Umklapp maximum at ~35 K. The total thermal conductivity $\kappa_{\perp c}$ is found to be higher than $\kappa_{\parallel c}$ in the entire temperature range, indicating a clear anisotropy in the conduction process in this compound. At room temperature, these values are $\kappa_{\perp c}$ = 0.916 W/m-K and $\kappa_{\parallel c}$ = 0.411 W/m-K. These values are comparable to the thermal conductivities of some of the well studied compounds like SnSe [20], doped PbTe [28], etc, thus making $In_2Te_5$ a promising thermoelectric material with a clear anisotropy in the heat conducting path. The anisotropy in thermal conductivity may be attributed to the layered structure [24] of the compound with very different lattice parameters. When the thermal gradient is applied perpendicular to the crystallographic *c*-axis (i.e., along the In-Te planes), the thermal currents encounter an easier path (higher thermal conductivity) as compared to the case when the thermal gradient is applied along the *c*-axis (i.e., perpendicular to the In-Te planes) due to the gaps between the layers which are connected only through weak van der Walls interactions [12, 24, 29].

The electronic part of the thermal conductivity ($\kappa_e$) is estimated from the Wiedmann- Franz law, $\kappa_e = L_0 \sigma T$, where $L_0$ is the Lorenz number 2.44 × $10^{-8}$ WΩ$K^{-2}$ and $\sigma$ is the electrical conductivity (obtained from the resistivity curves in Fig. 4). Figure 6(a) shows the calculated $\kappa_e$ values for the two configurations of the applied thermal gradient ($\kappa_{e\parallel c}$ for the parallel direction and $\kappa_{e\perp c}$ for the perpendicular direction) $\kappa_e$ values in both the directions increase with increasing temperature, consistent with the observed resistivity behavior. The lattice thermal conductivity ($\kappa_{l\parallel c}$ for the parallel direction and $\kappa_{l\perp c}$ for the perpendicular direction), after subtracting the electronic part from total thermal conductivity, is shown in Fig. 6(b). It is obvious from the figures that the electronic thermal conductivity is four orders of magnitude smaller than the phonon thermal conductivity and hence it can be argued that the phonon contributions towards $\kappa$ is dominant. This is further confirmed through the heat capacity measurements, as shown in Fig. 7, where it is shown that the heat capacity data can be fitted to the standard Debye model in the entire temperature range [30].

$$C_P = 9R\left(\frac{T}{\theta_D}\right)^3 \int_0^{x_D} dx \, \frac{x^4 \, e^x}{(e^x - 1)^2}$$

where $R$ is the gas constant and $\theta_D$ is the Debye temperature, $x_D = \left(\frac{\theta_D}{T}\right)$.

The temperature dependent power factor ($S^2\sigma$) along the two directions ($\parallel c$ and $\perp c$) are shown in Fig. 8(a). The power factor increases sharply above ~170 K along the $\parallel c$ direction and reaches a maximum value of 2.61 $\mu$W/K$^2$m at about 225 K as compared to that along the $\perp c$ axis which has a maximum value of only 1.45 $\mu$W/ K$^2$m. The temperature dependent figure of merit $zT$ along the two directions $zT_{\parallel c}$ and $zT_{\perp c}$ is shown in Fig. 8(b). $zT_{\parallel c}$ increases with temperature with a peak value of $1.6 \times 10^{-3}$ which is ~4 times larger than the $zT_{\perp c}$ ($4.1 \times 10^{-4}$) at about 270 K.

## 4. Conclusions

We have successfully grown the single crystals of $In_2Te_5$. By the thermopower measurements, we conclude that the majority charge carriers are holes. We have calculated the band gap through the Goldsmid and Sharp formula that is suggestive of the fact that $In_2Te_5$ could be considered as a narrow band semiconductor. We have observed through the direction dependent study that the lattice thermal conductivity along the $\parallel c$ direction is smaller by a factor of 2 as compared to the same along the $\perp c$ axis, indicating the anisotropy in this material. Similar anisotropic behavior is seen in thermopower and electrical resistivity of this compound. These, in turn, lead to the anisotropy in power factor and figure of merit. Our studies in $In_2Te_5$ single crystal reveals that, due to anisotropic behavior, a bulk material with layered structure, large unit cell and very low thermal conductivity can be used as a promising candidate in providing further scope to develop high efficiency thermoelectric materials. Even though the overall figure of merit values are small in this compound, suitable dopants can be used to investigate whether the electrical transport properties can be enhanced. Another aspect which can be improved to obtain higher values of figure of merit at higher temperatures is the suppression of the thermopower peak altogether or shifting it to higher temperatures.

**ACKNOWLEDGMENTS**

CVT would like to acknowledge the Department of Science and Technology for partial support through the project IR/S2/PU-10/2006.

**Figure Captions**

**Fig. 1.** Crystal structure of **(a)** $CsBi_4Te_6$ **(b)** $Bi_2Te_3$ **(c)** $Ag_2Te$ and **(d)** SnSe.

**Fig. 2.** **(a)** Powder X-ray diffraction pattern of $In_2Te_5$ with the details of Rietveld refinement. **(b)** X-ray diffraction pattern for a single crystal of $In_2Te_5$.

**Fig. 3.** **(a)** Crystal structure of $In_2Te_5$ viewed along the $b$-axis. **(b)** The HRTEM selected area electron diffraction pattern, indicating single crystalline nature. **(c)** HRTEM image of the $In_2Te_5$ single crystal. **(d)** Layered structure of monoclinic $In_2Te_5$ single crystal taken by FESEM.

**Fig. 4.** Variation of electrical resistivity ($\rho$) as a function of temperature ($T$) for $In_2Te_5$ single crystal parallel and perpendicular to $c$-axis.

**Fig. 5**. **(a)** Variation of Seebeck coefficient ($S$) and **(b)** total thermal conductivity ($\kappa$) as a function of temperature ($T$) for $In_2Te_5$ single crystal parallel and perpendicular to $c$-axis. Inset in the upper panel shows the schematic diagram of layers and the temperature gradient applied parallel and perpendicular to the $c$-direction.

**Fig. 6.** Variation of **(a)** $\kappa_e$ and **(b)** $\kappa_l$ with temperature ($T$) for $In_2Te_5$ single crystal parallel and perpendicular to $c$-axis.

**Fig. 7.** Measured specific heat ($Cp$) versus temperature ($T$) for $In_2Te_5$ single crystal. The estimated values from the Debye model is shown as a solid line.

**Fig. 8.** Variation of **(a)** power factor ($S^2\sigma$) and **(b)** figure of merit ($zT$) as a function of temperature ($T$) for $In_2Te_5$ single crystal parallel and perpendicular to $c$-axis.

**Table captions**

**TABLE I**. Summary of the crystallographic data and Rietveld refinement parameters for the single crystal of $In_2Te_5$.

**TABLE II.** Summary of composition and room temperature values of thermal conductivity ($\kappa$), resistivity ($\rho$), Seebeck coefficient ($S$), band gap ($E_g$) and figure of merit ($zT$) for $In_2Te_5$ single crystal in different directions.

# Figures

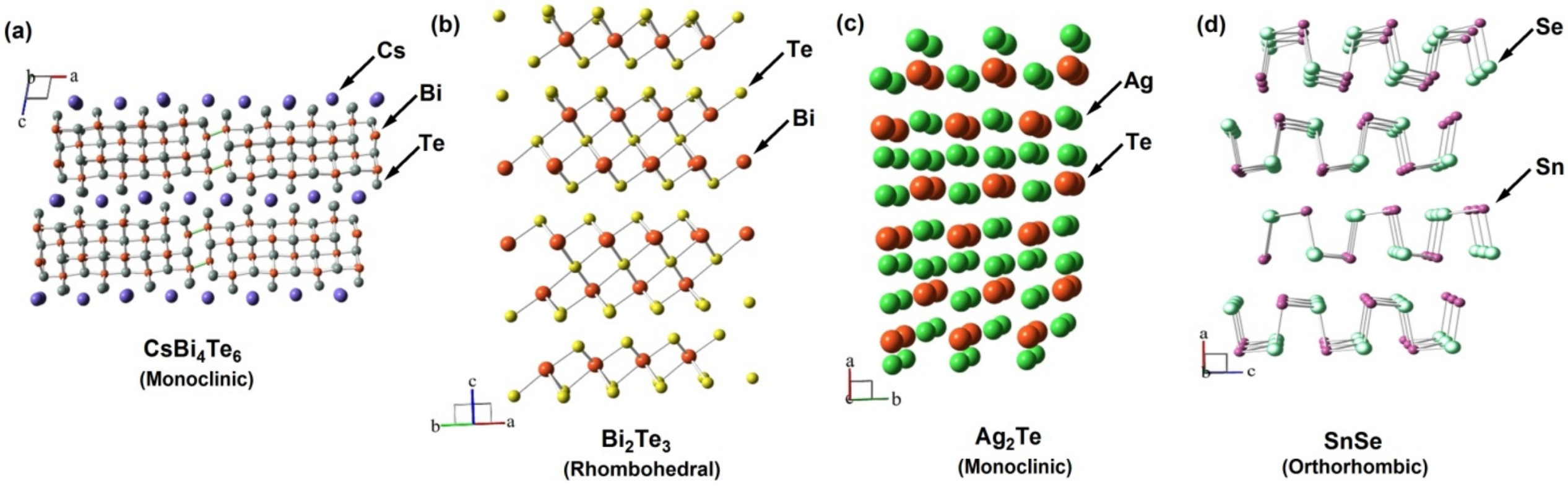

**Figure 1 (a), (b), (c), (d)**

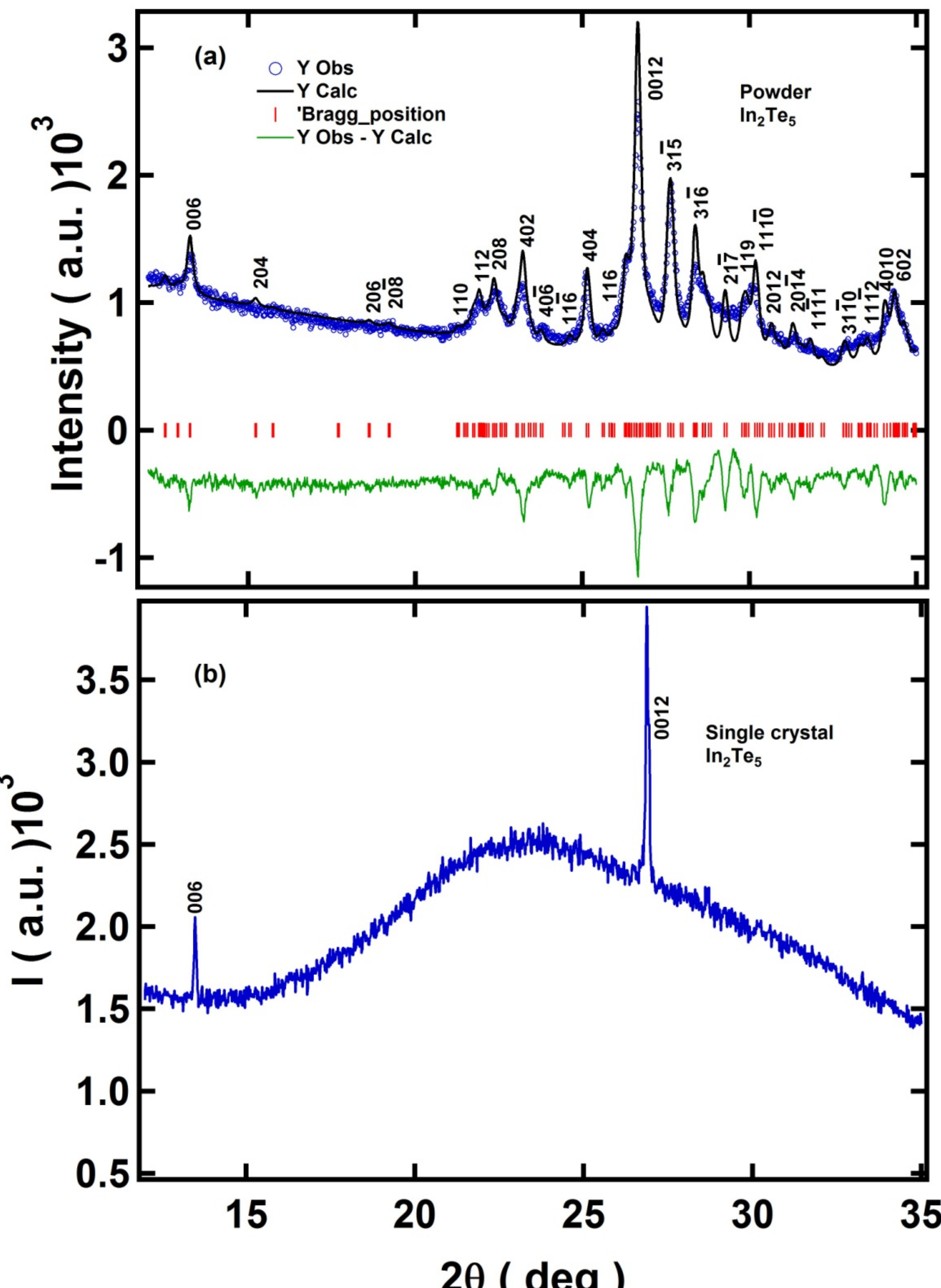

**Figure 2**

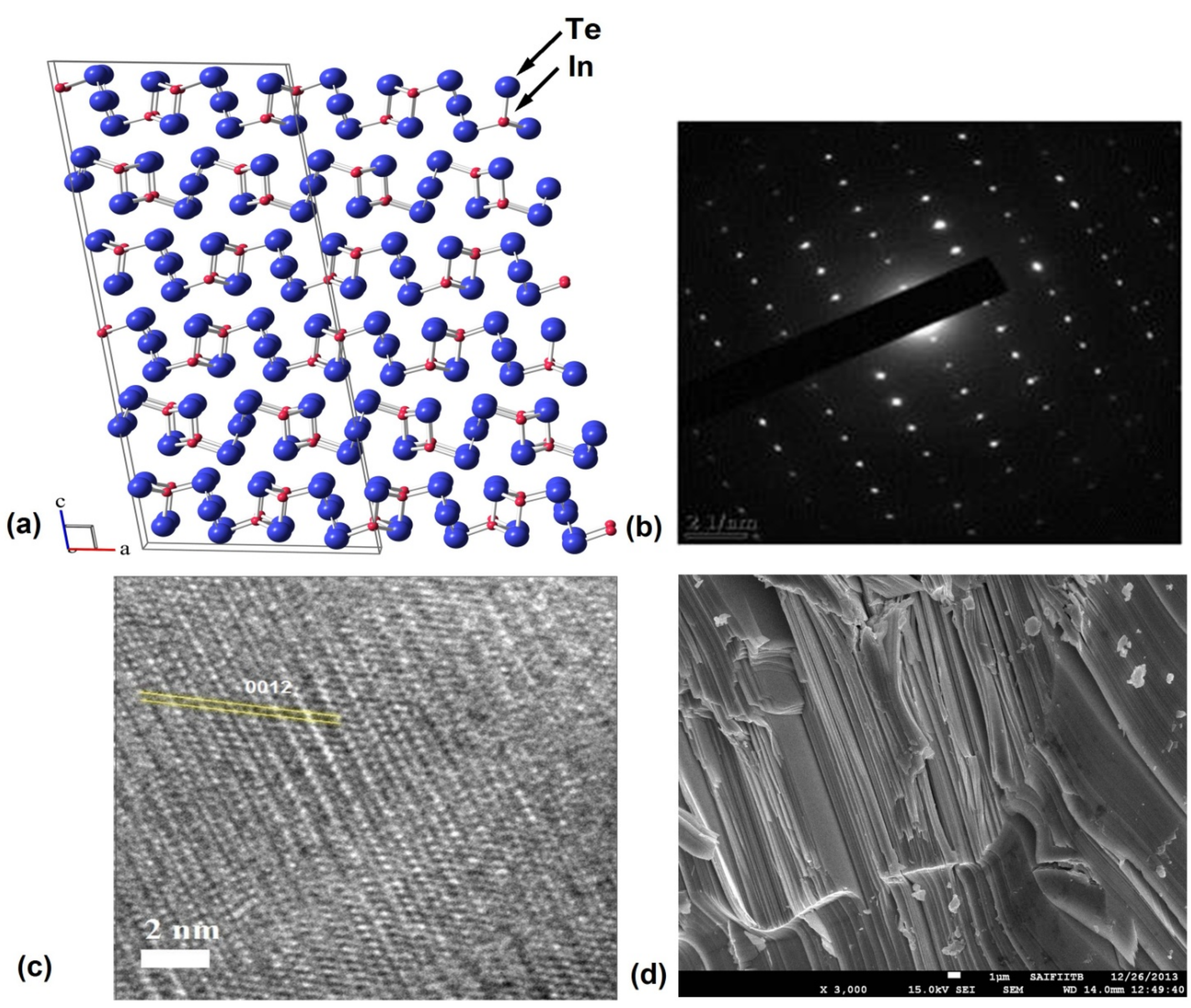

**Figure 3 (a), (b), (c), (d)**

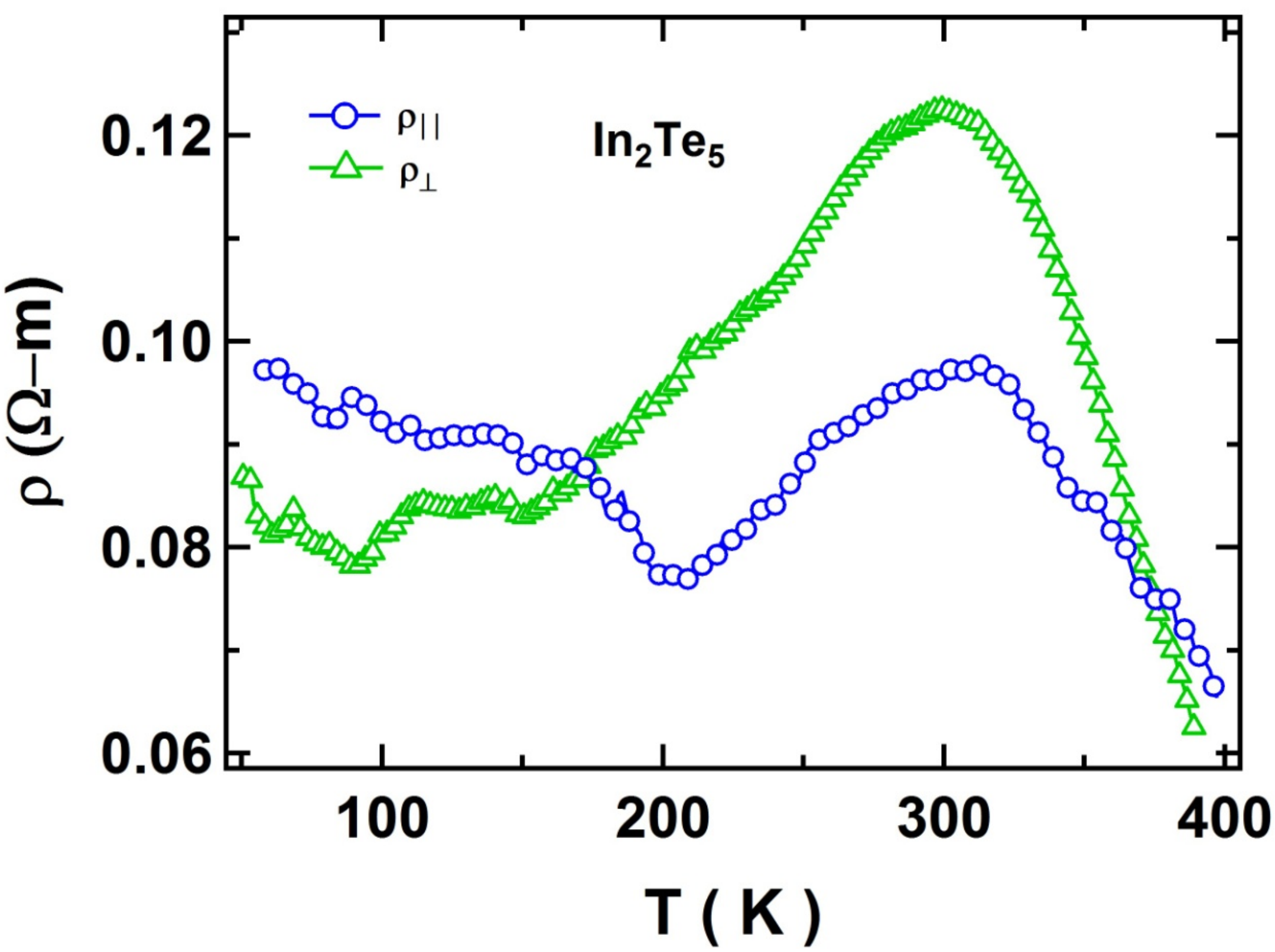

**Figure 4**

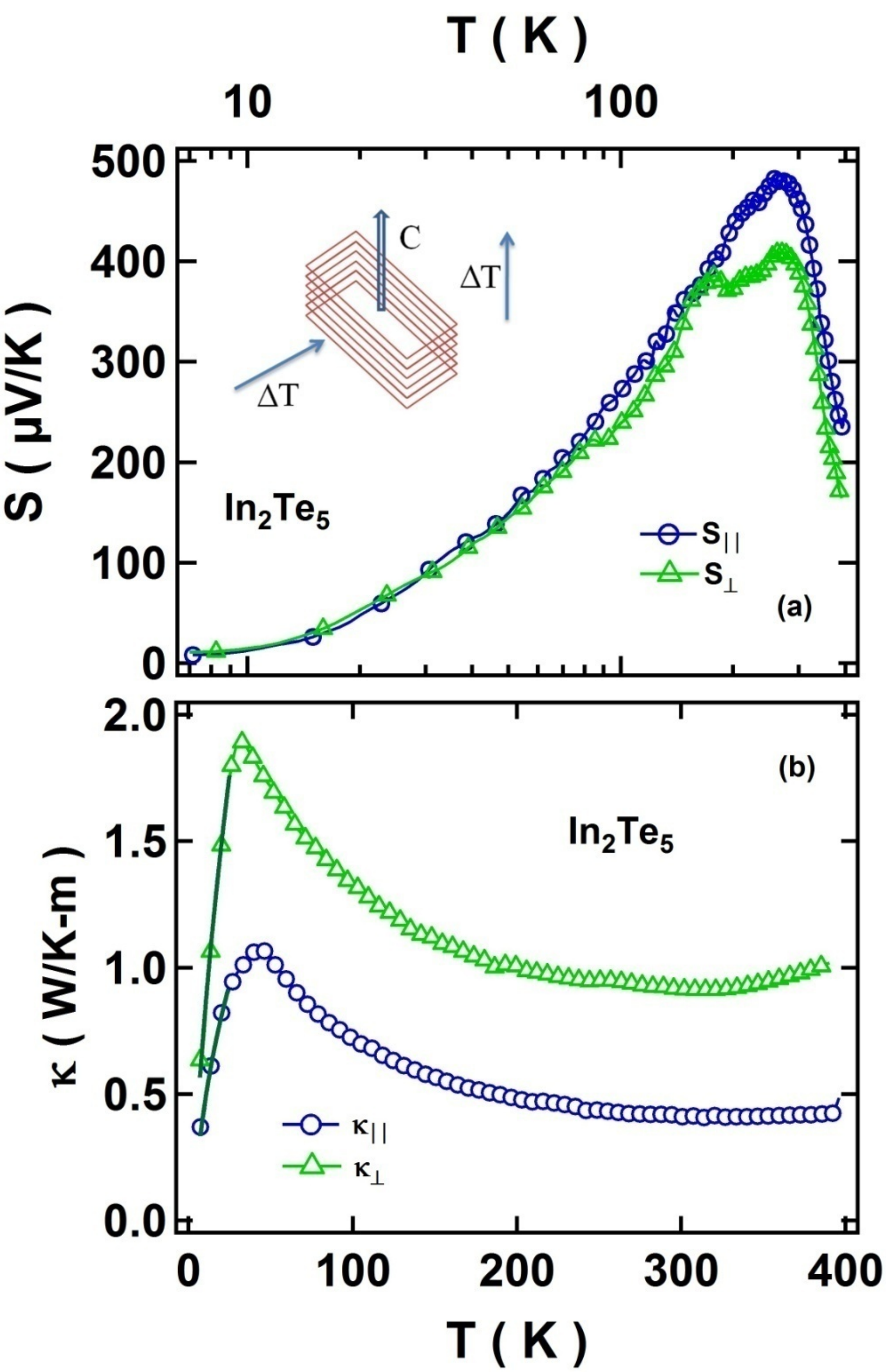

**Figure 5 (a), (b)**

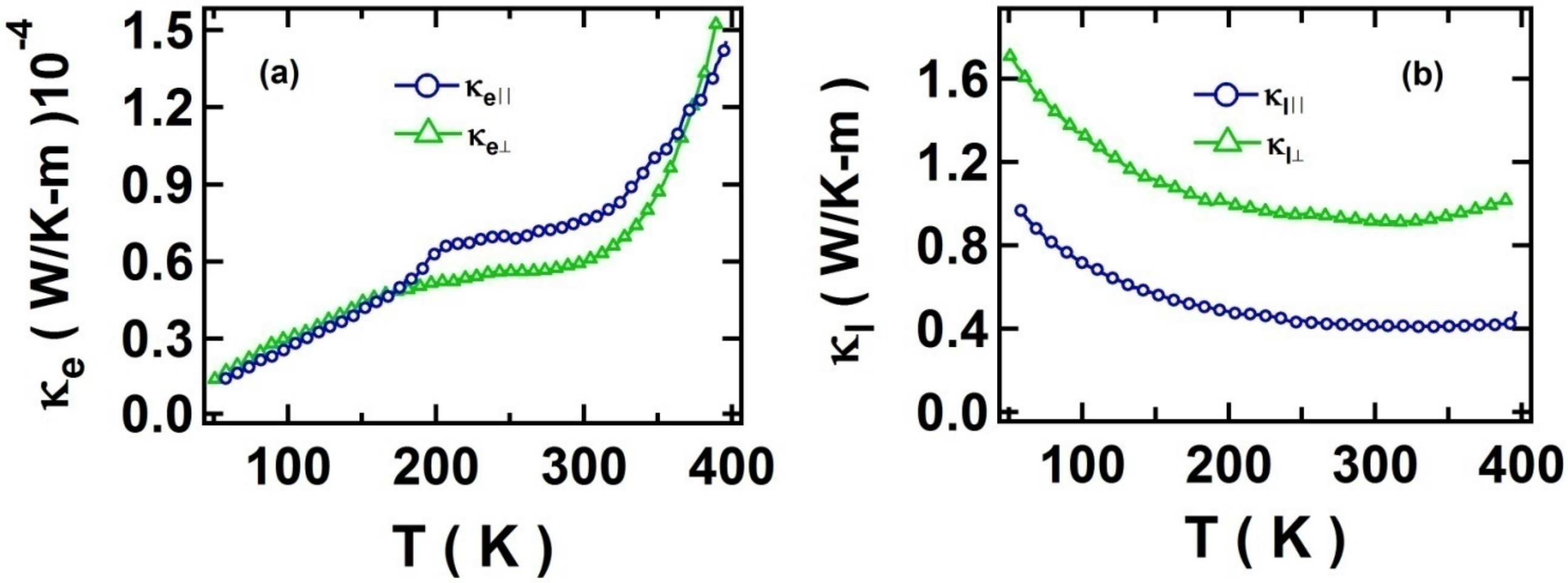

**Figure 6 (a), (b)**

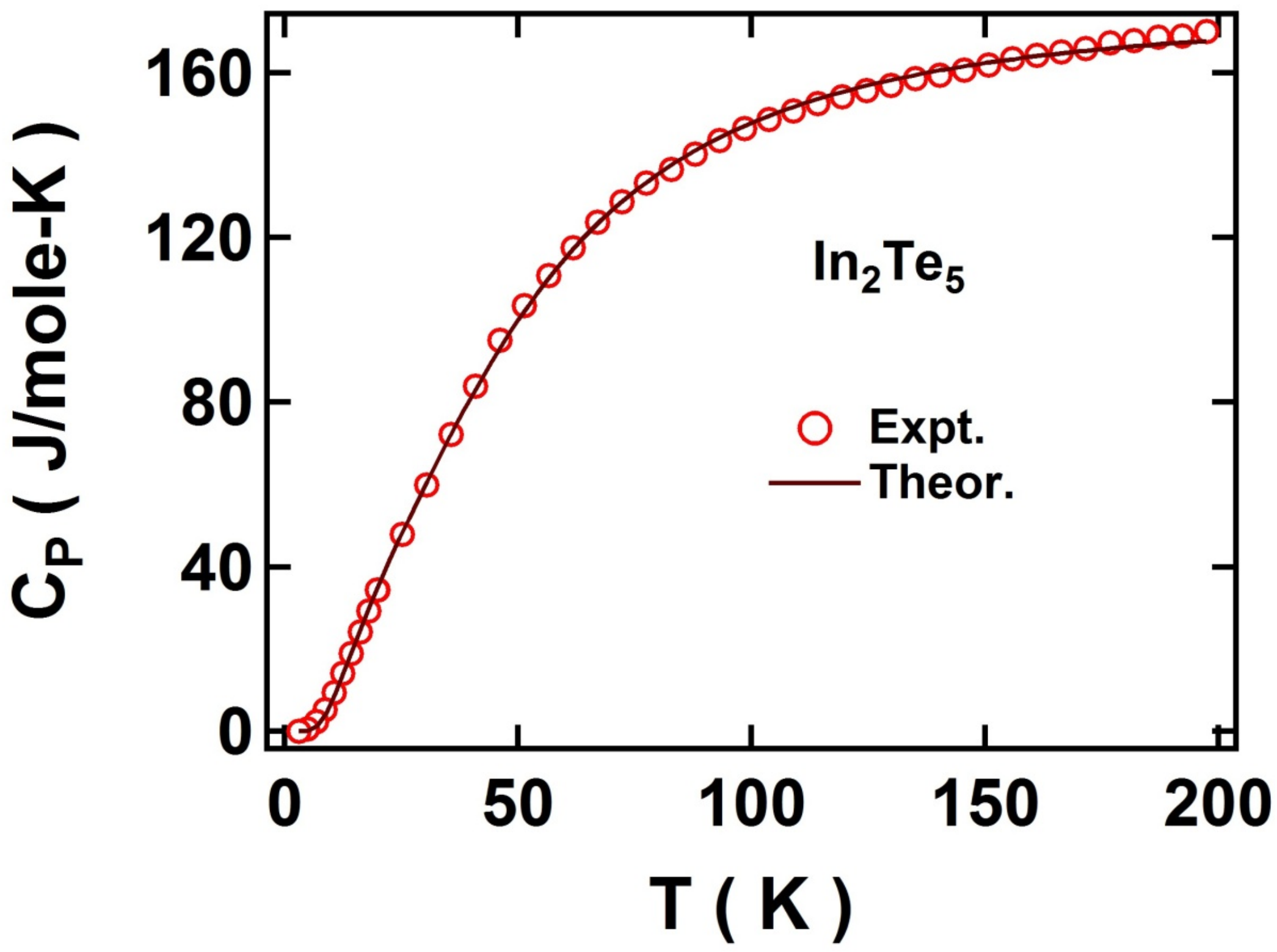

**Figure 7**

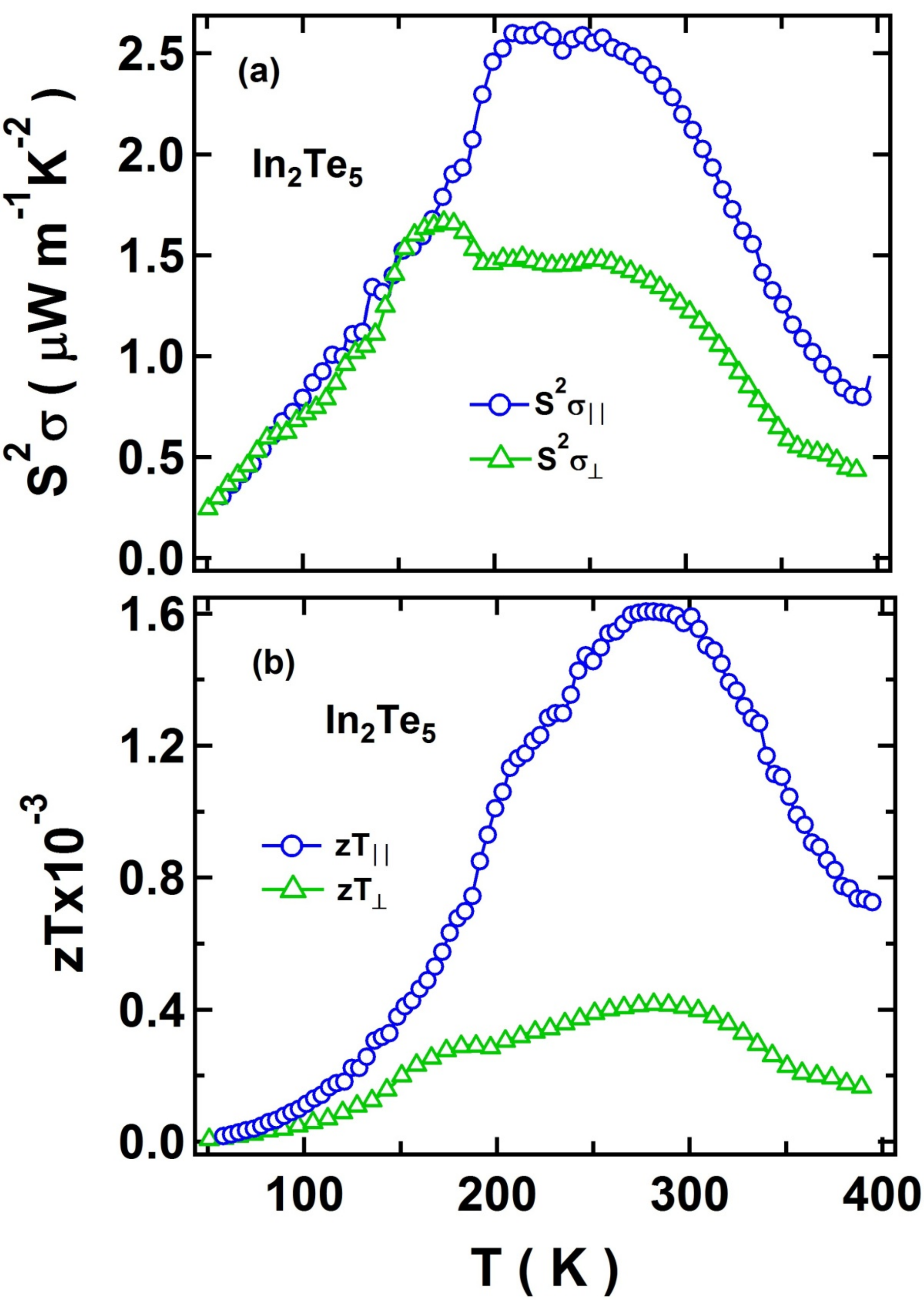

**Figure 8 (a), (b)**

**Table I**

| Sample | $In_2Te_5$ |
|---|---|
| Nature | Single crystal |
| Data | Powder XRD |
| Crystal Structure | Monoclinic |
| Space group | C12/c1 (15) |
| $a$ ( Å ) | 16.375 (5) |
| $b$ ( Å ) | 4.330 (1) |
| $c$ ( Å ) | 40.730 (1) |
| $V$ ( $Å^3$ ) | 2888 |
| $R_{exp}$ % | 3.418 |
| $R_{pro}$ % | 6.940 |
| $R_{wp}$ % | 9.225 |
| $\chi^2$ | 7.284 |

**Table II**

| Nominal composition | Average composition by EDAX | Direction | $\kappa$ (W/Km) | $\rho$ (Ωm) | $S$ (μv/K) | $E_g$ (eV) | $zT$ |
|---|---|---|---|---|---|---|---|
| $In_2Te_5$ | $In_{2.08}Te_{4.92}$ | $\|\|c$ | 0.41 | 0.096 | 460 | 0.27 | $1.6\times10^{-3}$ |
| | $In_{2.08}Te_{4.92}$ | $\perp c$ | 0.92 | 0.12 | 392 | 0.23 | $4.0\times10^{-4}$ |